\documentclass[epsfig,indentfirst,bm,12pt]{article}
\usepackage{CJK}
\usepackage{graphicx,amsmath,dcolumn,bm}
\begin{document}
\title{\Large{\bf{Atomic mass dependence of hadron production in semi-inclusive deep inelastic lepton-nucleus scattering }}
\thanks {Supported partially by  National Natural Science Foundation of China (11075044) and
Natural Science Foundation of Hebei Province (A2013209299).}}
\begin{CJK*}{GBK}{song}
\author{Li-Hua Song $^{1,3}$
 Na Liu  $^{1,4}$
  Chun-Gui Duan $^{1,2}$
\footnote{\tt{ E-mail:duancg$@$mail.hebtu.edu.cn}}}

\date{}

\maketitle

\end{CJK*}

\noindent {\small 1.Department of Physics, Hebei Normal
               University, Shijiazhuang 050024, China}\\
{\small 2.Hebei Advanced Thin Films Laboratory, Shijiazhuang 050024, China}\\
{\small 3.College of Science,  Hebei United University, Tangshan
063009, China}\\
{\small 4.College of Mathematics and Physics, Shijiazhuang
University of Economics, Shijiazhuang 050031, China}

\baselineskip 9mm
\begin{abstract}

Hadron production in lepton-nucleus deep inelastic scattering is
studied in a quark energy loss model.  The leading-order
computations for hadron multiplicity ratios are presented and
compared with the selected HERMES  pions production data with the
quark hadronization occurring outside the nucleus by means of the
hadron formation time. It is found that the  obtained energy loss
per unit length is $0.440 \pm 0.013$ GeV/fm for an outgoing quark by
the global fit. It is confirmed that the atomic mass number
dependence of hadron attenuation is  theoretically and
experimentally in good  agreement with the $A^{2/3}$ power law  for
quark hadronization occurring outside the nucleus.



\end{abstract}

\maketitle
\newpage
\vskip 0.5cm

\section*{1. Introduction}

The detailed understanding of the parton propagation and
hadronization processes in cold nuclear matter would greatly benefit
the study of the jet-quenching and parton energy loss phenomena
observed in ultra-relativistic heavy-ion collisions.  The
semi-inclusive deep inelastic scattering of lepton on nuclei  can
provide the essential information on the parton propagation and
hadronization processes in cold nuclear matter.  In the
semi-inclusive deep inelastic scattering on nuclei, a virtual photon
from the incident lepton is absorbed by a quark within a nucleus,
the highly virtual colored quark propagates over some distance
through the cold nuclear medium,  evolves subsequently into an
observed hadron. However,   the space-time evolution of the
hadronization process in cold nuclear matter are intrinsically
non-perturbative QCD processes. At present, the reliable QCD
calculations of parton hadronization  cannot yet be performed.
Therefore, the investigation into this process from the partons
produced in the elementary interaction to the observed hadrons on
the phenomenological level is of basic importance for development of
the theoretical research[1].

Two classes of theoretical phenomenological models were proposed to
describe the experimental data from the semi-inclusive deep
inelastic scattering of lepton on the nucleus. The experimental data
are usually presented in terms of the multiplicity ratio $R^h_M$,
which is defined as the ratio of the number of hadrons h produced
per deep-inelastic scattering  event on a nuclear target with mass
number A to that for a deuterium  target[2-9]. The first class are
the absorption-type models[10-15], which commonly presume that $1-
R^h_M \sim A^{1/3}$. Another class of models are the parton energy
loss models[16-19], which focus on the parton energy loss that the
struck quark experiences in the nuclear environment, and assume that
the hadron is formed outside the nucleus without hadron absorption
in the nuclear medium. The parton energy loss models predict that
$1- R^h_M \sim A^{2/3}$. It is obvious that the different physical
mechanism in two classes of phenomenological models directly leads
to the different prediction on atomic mass dependence of hadron
production in deep inelastic scattering on nuclei.  However, the
recent researches indicate that atomic mass dependence of hadron
production is far from being expected by the absorption and parton
energy loss models.  The authors of Ref. [20,21] find that the mass
number dependence of the hadron attenuation $1- R^h_M$ obeys a
$A^{2/3}$ law (broken at $A \geq 80$) in both energy loss and
absorption models.  The observed approximate $A^{2/3}$  scaling of
experimental data for light nuclei cannot be used as a proof of the
energy loss mechanism. In addition, the study on the attenuation of
hadron production by using realistic matter distributions[22] shows
that the mass number dependence for a pure partonic (absorption)
mechanism is more complicated than a simple $A^{2/3}(A^{1/3})$
behavior. For this reason, it is hoped that the precise experimental
measurement on atomic mass dependence of hadron production would
allow us to clearly know the physical mechanism on the hadronization
process.

In our preceding article[23], we have calculated the nuclear
modifications of hadron production in semi-inclusive deep inelastic
scattering  in a parton energy loss model. By means of the short
hadron formation time, the relevant data with quark hadronization
occurring outside the nucleus are picked out from HERMES
experimental results[7] on the one-dimensional dependence of the
multiplicity ratio $R^h_M$ as a function of the energy fraction $z$
of the virtual photon carried away by the hadron. Our theoretical
results show that the nuclear effects on parton distribution
functions can be neglected. We find that the theoretical results
considering the nuclear modification of fragmentation functions due
to quark energy loss are in good agreement with the selected
experimental data. The experimental data on the hadron multiplicity
ratio do not distinguish between the linear and quadratic dependence
of quark energy loss. The obtained energy loss per unit length is
$0.38 \pm 0.03$ GeV/fm for an outgoing quark by the global fit.
However, it is worth to mention that the selected experimental data
only involve the multiplicity ratio on helium and neon targets
relative to deuterium.

In this paper we extend the proposed quark energy loss model to
study the atomic mass dependence of the nuclear attenuation $1-
R^h_M $ because whatever the physical mechanism, the atomic mass
dependence will be an important ingredient. Although the new
HERMES[8] and CLAS[9] data are reported on the multiplicity ratios
for three nucleus targets relative to deuterium, in order to explore
the  atomic mass dependence of the nuclear attenuation, we employ
the so called two dimensional data from HERMES on the multiplicities
for the production of pions on helium, neon, krypton, and xenon
targets relative to those for deuterium[7]. Two dimensional data
mean that the multiplicity ratio is presented in a fine binning in
one variable and a coarser binning in another variable. The other
variables are integrated over within the acceptance of the
experiment.  The experimental data with quark hadronization
occurring outside the nucleus are selected by means of the  hadron
formation time. It is hoped to gain new knowledge about the atomic
mass dependence of the nuclear attenuation.

The remainder of the paper is organized as follows. The brief
formalism for the hadron  multiplicity in semi-inclusive deep
inelastic scattering on the nucleus and nuclear modification of the
fragmentation functions owing to quark energy loss are described in
Section 2. Then, the numerical computations for $R^h_M $ and the
A-dependence of $1- R^h_M $ are presented and compared with
experimental data. Finally, a summary is presented.

\section*{2. The hadron  multiplicity in semi-inclusive deep inelastic
scattering on nuclei }

At leading order in perturbative QCD, the hadron multiplicity can be
obtained from normalizing the semi-inclusive deep inelastic lepton
nucleus scattering yield $N^{h}_{A}$ to the deep inelastic
scattering yield $N^{DIS}_{A}$,
\begin{equation}
\frac{1}{N^{DIS}_{A}}\frac{dN^{h}_{A}}{dzd
\nu}=\frac{1}{\sigma^{lA}}\int
dx\sum_{f}e^{2}_{f}q^{A}_{f}(x,Q^{2})\frac{d\sigma^{lq}}{dxd\nu}D^{A}_{f|
h}(z,Q^{2}),
\end{equation}
\begin{equation}
\sigma^{lA}=\int
dx\sum_{f}e^{2}_{f}q^{A}_{f}(x,Q^{2})\frac{d\sigma^{lq}}{dxd\nu},
\end{equation}
\begin{equation}
\frac{d\sigma^{lq}}{dxd\nu}=Mx\frac{4\pi\alpha_{s}^{2}}{Q^{4}}[1+(1-y)^{2}].
\end{equation}
In the above equation, $\nu$ is the virtual photon energy, $e_f$ is
the charge of the quark with flavor $f$, $q^{A}_{f}(x,Q^{2})$ is the
nuclear quark distribution function with Bjorken variable $x$ and
photon virtuality $Q^{2}$, ${d\sigma^{lq}}/{dxd\nu}$ is the
differential cross section for lepton-quark scattering at leading
order,  $D^{A}_{f| h}(z,Q^{2})$ is the nuclear modified
fragmentation function of a quark of flavour $f$ into a hadron $h$,
and $ \alpha _{s}$ and $y$ are the fine structure constant and the
fraction of the incident lepton energy transferred to the target,
respectively.

In a similar way to the  parton propagation in nuclear Drell-Yan
process[24,25], the struck quark by the virtual photon can lose its
energy owing to multiple scattering from other quarks and gluon
radiation while traversing through the nucleus. In our previous
article[23],  the quark energy loss  is written as
\begin{equation}
\Delta E=\alpha <L>_{_{A}},
\end{equation}
which is often referred to as the linear quark energy loss.  Here,
$\alpha$ is the fitted parameter that can be extracted from
experimental data. As for quark hadronization occurring outside the
nucleus, $<L>_{A}$ is equal to $3/4(1.12A^{1/3})$fm.  With
considering the quark energy loss in nucleus target, the quark
energy fragmenting into a hadron shifts from $E_{q}=\nu$ to
$E'_{q}=\nu-\Delta E$, which results in a rescaling of the energy
fraction of the produced hadron:
\begin{equation}
z=\frac{E_{h}}{\nu}  \longrightarrow  z'=\frac{E_{h}}{\nu-\Delta
E},
\end{equation}
where $E_{h}$ and $\Delta E$ are, respectively, the measured hadron
energy and the quark energy loss in the nuclear medium. In view of
the rescaled energy fraction of the produced hadron, the
fragmentation function in the nuclear medium[23]
\begin{equation}
D^{A}_{f|h}(z,Q^{2})=D_{f|h}(z',Q^{2}),
\end{equation}
where $D_{f|h}$ is the standard (vacuum) fragmentation function of a
quark of flavour $f$ into a hadron $h$.  Therefore, excluding the
influence of nuclear absorption, and considering only the case where
hadrons are produced outside the nucleus, the hadron multiplicity
can be expressed as
\begin{equation}
\frac{1}{N^{DIS}_{A}}\frac{dN^{h}_{A}}{dzd
\nu}=\frac{1}{\sigma^{lA}}\int
dx\sum_{f}e^{2}_{f}q^{A}_{f}(x,Q^{2})\frac{d\sigma^{lq}}{dxd\nu}D_{f|h}(z',Q^{2}).
\end{equation}

\section*{3. Results and discussion}

The HERMES Collaboration at the DESY laboratory  performed a series
of semi-inclusive deep-inelastic scattering measurements on
deuterium, helium, neon, krypton, and xenon targets  in order to
study hadronization by using a 27.6 GeV positron or electron
beam[7]. The hadron multiplicity ratios  on  pions two-dimensional
distributions  were presented.

In the following calculation, we use the two-dimensional data on the
multiplicity ratio $R^h_M$ for pions produced on helium, neon,
krypton and xenon nuclei, in three z slices as a function of $\nu$,
and in three $\nu$ slices as a function of $z$. We pick out the
experimental data with quark hadronization occurring outside the
nucleus by means of the hadron formation time
$t=z^{0.35}(1-z)\nu/\kappa$ ($\kappa = 1GeV/fm$)[7,26], which is
defined as the time between the moment that the quark is struck by
the virtual photon and the moment that the prehadron is formed. If
$t>3/4R_A$, hadronization occurs outside the nucleus. Otherwise,
hadrons are produced inside the nucleus. Using the criterion, the
number of points in selected experimental data is 48 for
$R^{\pi}_{M}(\nu)$ in two $z$ regions of $0.2< z < 0.4$ and $0.4 < z
< 0.7$.  For the multiplicity ratio $R^{\pi}_{M}$ as a function of
$z$, there are  57 data points in three $\nu$ regions of 6 $< \nu <$
12 GeV , 12 $< \nu <$ 17 GeV and 17 $< \nu <$ 23.5 GeV.

As for the cases with quark hadronization occuring outside the
nucleus, we compute at leading order  the hadron multiplicity ratios
$R^{\pi}_{M}$,
\begin{equation}
R^{\pi}_{M}[\nu(z)]=\int
\frac{1}{N^{DIS}_{A}}\frac{dN^{h}_{A}(\nu,z)}{dzd\nu}dz(\nu)
\Bigg/\int
\frac{1}{N^{DIS}_{D}}\frac{dN^{h}_{D}(\nu,z)}{dzd\nu}dz(\nu),
\end{equation}
for the production of pions by using the CTEQ6L parton density in
the proton[27]  together with the vacuum fragmentation functions
[28], meanwhile taking account of the quark energy loss  in cold
nuclear matter. In our calculation, the integral range is determined
according to the relative experimental kinematic region. Then,  the
obtained  hadron multiplicity ratios $R^{\pi}_{M}$ are compared with
the experimental values for calculating
\begin{equation}
\chi^{2}=\sum_{i}^{m} \Bigg[\frac{R^{\pi, data}_{M, i}-R^{\pi,
theo}_{M, i}}{\sigma^{err}_{i}} \Bigg]^2,
\end{equation}
where $R^{\pi, data}_{M, i}$ and $R^{\pi, theo}_{M, i}$ indicate
separately the experimental data and theoretical values of the
hadron multiplicity ratio $R^{\pi}_{M}$, the experimental error is
given by systematic and statistical errors as
$(\sigma^{err}_{i})^2=(\sigma^{syst}_{i})^2+(\sigma^{stat}_{i})^2$.
Together with the CERN subroutine MINUIT[29], the optimum parameter
in quark energy loss expression is obtained by minimizing $\chi^2$.
One standard deviation of the optimum parameter in quark energy loss
expressions corresponds to an increase of $\chi^{2}$ by 1 unit from
its minimum $\chi^{2}_{min}$.

For all of the selected experimental data on the hadron multiplicity
ratios $R^{\pi}_{M}$ as a function of $z$ and $\nu$, our analysis
has in total 105 data points, and 4 nuclei from helium up to xenon.
The global fit of all data makes $\alpha =0.440 \pm 0.013 $ with the
relative uncertainty $\delta \alpha / \alpha \simeq 3\%$ and
$\chi^2/ndf = 1.02$. It is worthy of note that the value of energy
loss per unit length $\alpha$ is a little bit  bigger  than that in
our previous article[23]. But, we think the present value is more
realistic because of the precise two dimensional experimental data
on the hadron multiplicity ratios.

To demonstrate intuitively the quark energy loss  effect on hadron
multiplicity ratio $R^{\pi}_{M}$, the experimental data selected for
the present analysis are respectively compared with our theoretical
predictions on $R^{\pi}_{M}(\nu)$ in two $z$ domains in  figure 1
and $R^{\pi}_{M}(z)$ in three $\nu$ domains in  figure 2 for
helium(solid line), neon(dashed line), krypton(dotted  line) and
xenon(dash dot line) nuclear targets.  It is shown that the
theoretical results including the nuclear modification of
fragmentation functions owing to quark energy loss are in good
agreement with the experimental data on the hadron multiplicity
ratios.

\begin{figure}[t,m,b]
\centering
\includegraphics*[width=15cm,height=9cm]{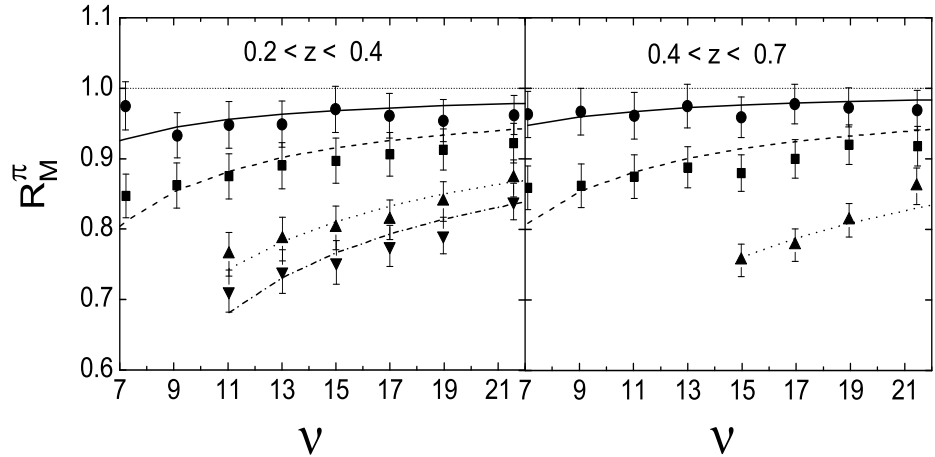}
\vspace{-1.1cm} \caption{The calculated multiplicity ratios
$R^{\pi}_{M}(\nu)$ in two $z$ regions for pions production on He
(solid line), Ne (dashed line), Kr (dotted  line) and Xe (dash dot
line) nuclei from the linear quark energy loss. The HERMES data[7]
on He (solid circles), Ne (filled boxes), Kr (closed triangles) and
Xe (inverted triangles) are shown with  the total uncertainty
(statistical plus systematic, added quadratically).}
\end{figure}

\begin{figure}[t,m,b]
\centering
\includegraphics*[width=16cm,height=7.0cm]{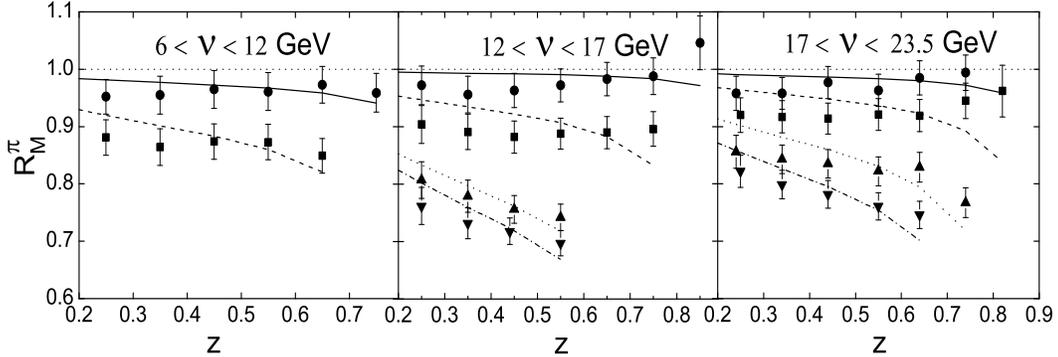}
\vspace{-1.0cm} \caption{The calculated multiplicity ratios
$R^{\pi}_{M}(z)$ in three $\nu$ regions for pions production on He
(solid line), Ne (dashed line), Kr (dotted  line) and Xe (dash dot
line) nuclei from the linear quark energy loss. The other comments
are the same as those in Fig.1.}
\end{figure}

In order to investigate the atomic mass dependence of hadron
production in semi-inclusive deep inelastic lepton-nucleus
scattering, the hadron attenuation $1-R^{\pi}_{M}$ is presumed in
terms of a power law:
\begin{equation}
1-R^{\pi}_{M}=cA^{\kappa}.
\end{equation}
In general, the coefficient $c$ and exponent $\kappa$ both depend on
the kinematic variable $z(\nu)$ and the atomic mass number $A$. The
optimal parameters $c$ and $\kappa$ in  the power law can be
determined by chi-square minimization, i.e., the $\chi^{2}$ merit
function
\begin{equation}
\chi^{2}(c, \kappa)=\sum_{i}^{m}
\Bigg[\frac{(1-R^{\pi}_{M})(A_i)-cA^{\kappa}_i}{\sigma^{err}_{i}}
\Bigg]^2,
\end{equation}
is minimized with respect to $c$ and $\kappa$. Here
$\sigma^{err}_{i}$ is the uncertainty of the theoretical or the
experimental points, respectively. For the case of the theoretical
computations,  we evaluate the uncertainty of $R^{\pi}_{M}[\nu(z),
\alpha]$ with respect to the optimized parameter $\alpha$  by using
Hessian method and assuming linear error propagation:
\begin{equation}
\delta R^{\pi}_{M}[\nu(z), \alpha]=\triangle\chi^{2} \frac{\partial
R^{\pi}_{M}[\nu(z), \alpha]}{\partial \alpha}\Bigg[ \frac{\partial^2
R^{\pi}_{M}[\nu(z), \alpha]}{\partial^2 \alpha} \Bigg]^{-1/2}.
\end{equation}
In our estimation, the $\triangle\chi^{2}$ value is obtained by the
following procedure. The confidence level $P$ could be chosen as the
one-$\sigma$-error range of the normal distribution. For one
parameter, $P=0.6826$ is obtained with $\triangle\chi^{2}=1$.

\begin{table}[t,m,b]
\caption{The  coefficient $c$ and exponent $\kappa$  with their
uncertainties for the fit $1-R^{\pi}_{M}=cA^{\kappa}$ for pions
production at fixed $\nu$ and $z$ bins, both for the selected
experimental data and our model calculation. The nuclei included in
the fits are helium, neon, krypton and xenon nuclei.}
\begin{center}
\begin{tabular}{p{3cm}p{3cm}p{3cm}p{3cm}c}\hline
          & \multicolumn{2}{c}{Experiment}& \multicolumn{2}{c}{Theory}
          \\ \cline{2-3} \cline{4-5}
         $\nu$  & $ c[10^{-2}]$ &$\kappa $& $ c[10^{-2}]$ &$\kappa $\\
         \hline
         & \multicolumn{4}{c}{0.2$< z <$ 0.4} \\ \cline{2-5}
       11.04& $4.26\pm3.93 $ & $0.39\pm0.20 $& $1.49\pm0.91$ & $0.60\pm0.13$\\
       13.01 &$3.87\pm4.01$ &$0.39\pm0.22 $& $1.23\pm0.75$& $0.61\pm0.13$\\
       15.00  & $2.98\pm3.59 $& $0.43\pm0.25$& $1.06\pm0.64$& $0.61\pm0.13 $\\
       16.99 &$3.16\pm3.73 $&$0.40\pm0.25$& $0.93\pm0.55$ & $0.61\pm0.13$\\
      18.98  & $3.16\pm3.76 $&$0.38\pm0.25$& $0.84\pm0.49$ & $0.61\pm0.13 $\\
      21.57  &$3.01\pm3.91$&$0.34\pm0.27$& $0.76\pm0.43$& $0.61\pm0.12 $\\
\hline
      $z$ & \multicolumn{4}{c}{12 $< \nu <$ 17 GeV} \\ \cline{1-5}
        0.25 & $2.74\pm3.59 $ & $0.44\pm0.27 $& $1.00\pm0.47 $ & $0.63\pm0.10 $\\
       0.35 &$3.54\pm3.60 $&$0.41\pm0.22 $& $1.25\pm0.59 $ & $0.62\pm0.10$\\
       0.55  &$3.21\pm2.93 $&$0.46\pm0.19 $& $1.81\pm0.84 $ & $0.61\pm0.10 $\\ \hline
     & \multicolumn{4}{c}{17 $< \nu <$ 23.5 GeV} \\ \cline{2-5}
      0.34  & $2.93\pm3.55 $ & $0.39\pm0.26 $& $0.97\pm0.70 $ & $0.60\pm0.15$\\
       0.44 &$2.32\pm3.14$&$0.45\pm0.28 $& $1.17\pm0.84$& $0.60\pm0.15 $\\
       0.55  & $2.27\pm3.27 $&$0.48\pm0.30 $& $1.41\pm1.03 $ & $0.59\pm0.16 $\\
       0.64  &$1.63\pm3.10$ &$0.55\pm0.37 $& $1.77\pm1.27$ & $0.59\pm0.15 $\\
\hline
\end{tabular}
\end{center}
\end{table}

The best-fit  coefficient $c$ and exponent $\kappa$  with their
uncertainties for the fit $1-R^{\pi}_{M}=cA^{\kappa}$ are
 summarized in  Table 1 at fixed $\nu$ bins in the
region $0.2 < z < 0.4$ and  at fixed $z$ bins in two regions of  $12
< \nu < 17 $ GeV and $17 < \nu < 23.5 $ GeV both from the
experimental data[7] and our model calculation on the hadron
attenuation $1-R^{\pi}_{M}$ for pions production on helium, neon,
krypton and xenon nuclei. As can be seen from the Table 1, the
relative uncertainties of  coefficient $c$, $\delta c/c$, are very
large for  the selected experimental data and our model calculation.
The large relative uncertainties $\delta c/c$ manifest that the
coefficient $c$ does not determined well. In addition, the values of
exponent $\kappa$ obtained from our model are overall larger than
those from the experimental data while the values of coefficient $c$
given by our model prediction are in general smaller than those from
the experimental data. The fact indicates that the exponent $\kappa$
and coefficient $c$ are strongly correlated.

\begin{table}[t,m,b]
\caption{The  coefficient c and  $\chi^2/ndf$  by the fit
$1-R^{\pi}_{M}=cA^{2/3}$ for pions production at fixed $\nu$ and $z$
bins, both for the selected experimental data and our model
calculation. The nuclei included in the fits are helium, neon,
krypton and xenon nuclei.}
\begin{center}
\begin{tabular}{p{3cm}p{3cm}p{3cm}p{3cm}c}\hline
          & \multicolumn{2}{c}{Experiment}& \multicolumn{2}{c}{Theory}
          \\ \cline{2-3} \cline{4-5}
         $\nu$  & $ c[10^{-2}]$ &$\chi^{2}/ndf$& $ c[10^{-2}]$ &$\chi^{2}/ndf$\\
         \hline
         & \multicolumn{4}{c}{0.2$< z <$ 0.4} \\ \cline{2-5}
         11.04& $1.19\pm0.20 $ & $1.83 $& $1.12\pm0.25$ & $0.20$\\
       13.01 &$1.08\pm0.21$ &$1.37$& $0.94\pm0.21$& $0.20$\\
       15.00  & $1.01\pm0.21 $& $1.07 $& $0.82\pm0.19$& $0.20$\\
       16.99 &$0.93\pm0.19$&$1.06$& $0.72\pm0.16$ & $0.20$\\
      18.98  & $0.84\pm0.18$&$1.11$& $0.65\pm0.15$ & $0.20$\\
      21.57  &$0.66\pm0.18$&$1.11$& $0.59\pm0.13$& $0.21$\\ \hline
         $z$ & \multicolumn{4}{c}{12 $< \nu <$ 17 GeV} \\ \cline{1-5}
        0.25 & $0.97\pm0.21$ & $0.85$& $0.83\pm0.11$ & $0.34$\\
       0.35 &$1.10\pm0.18$&$1.37$& $1.00\pm0.14$ & $0.38$\\
       0.55  &$1.25\pm0.16$&$1.61$& $1.39\pm0.19$ & $0.43$\\ \hline
     & \multicolumn{4}{c}{17 $< \nu <$ 23.5 GeV} \\ \cline{2-5}
      0.34  & $0.81\pm0.17$ & $1.03$& $0.72\pm0.14$ & $0.17$\\
       0.44 &$0.87\pm0.17$&$0.86$& $0.85\pm0.17$& $0.18$\\
       0.55  & $0.94\pm0.18$&$0.49$& $1.01\pm0.20$ & $0.18$\\
       0.64  &$0.96\pm0.20$ &$0.50$& $1.25\pm0.23$ & $0.19$\\
\hline
\end{tabular}
\end{center}
\end{table}

In  consideration of the strong correlation between the exponent
$\kappa$ and coefficient $c$, we further assume the hadron
attenuation $1-R^{\pi}_{M}$ in terms of the power law:
\begin{equation}
1-R^{\pi}_{M}=cA^{2/3}.
\end{equation}
Here the coefficient $c$ is a free parameter. The best-fit parameter
$c$ can be pinned down by chi-square minimization. Table 2 summarize
the  coefficient c and  $\chi^2/ndf$  by the fit
$1-R^{\pi}_{M}=cA^{2/3}$ for pions production at fixed $\nu$ and $z$
values from the selected experimental data and our model
calculation. It is shown that the values of coefficient $c$ increase
(decrease) with increasing values of $z (\nu)$. Regarding the
numerical results from our model, the atomic mass number dependence
of hadron attenuation is shown to agree well with the $A^{2/3}$
power law. In addition, the computed results from the selected
experimental data with the quarks hadronization occurring outside
the nucleus, are overall consistent with  the theoretical prediction
on $1- R^{\pi}_M \sim A^{2/3}$ from the parton energy loss model.
Therefore, we can conclude apparently that with marked difference
from the relative research results[20-22], our model and the
selected experimental data from HERMES are conditionally in support
of the power law $1-R^{\pi}_{M} \sim A^{2/3}$.

\section*{4. Summary }

Hadron production in lepton-nucleus deep inelastic scattering is
intrinsically non-perturbative QCD process without the  reliable QCD
calculations. In the proposed quark energy loss model, the
experimental data with the quarks hadronization occurring outside
the nucleus are selected by means of the hadron formation time. We
perform a leading order phenomenological analysis on the hadron
multiplicity ratio, and compare with the selected HERMES
experimental results for pions produced on helium, neon, krypton and
xenon nuclei. Our results show that the theoretical expectations
with the fragmentation functions modified due to quark energy loss
are in good agreement with the experimental data.  We obtain the
outgoing quark energy loss per unit path length $\alpha =0.440 \pm
0.013 $ GeV/fm from the global fit of all selected data. As for
quark hadronization occurring outside the nucleus, it is
demonstrated quantitatively that the mass number dependence of the
hadron attenuation $1-R^{h}_{M}$ agrees well with the $A^{2/3}$
power law from both  the selected experimental data and our model
calculation. Our study manifests that the process of hadronization
in cold nuclear matter is most probably a combination of the parton
energy loss and hadronic absorption.

\vskip 1cm

\end{document}